\documentstyle[aps]{revtex}
 
\newcommand{\putsmall}[1]{ \begin{center} 
\vbox to 100bp{ \vfil \hbox to 330bp{
\includegraphics{#1}
\hfil }} \end{center} }
\font\eightrm=cmr8

\begin{document}
\draft

\twocolumn

\noindent
{\bf Comment on ``Lyapunov Exponent of a Many Body System
and Its Transport Coefficients''}
 
\noindent
In a recent Letter, Barnett, Tajima, Nishihara, Ueshima and Furukawa 
\cite{barn1} obtained a theoretical expression for the maximum
Lyapunov exponent $\lambda_1$ of a dilute gas. They conclude that
$\lambda_1$ is proportional to the cube root of the self-diffusion
coefficient $D$, independent of the range of the
interaction potential.  They validate their conjecture with
numerical data for a dense one-component plasma, a system with 
long-range forces.
We claim that their result is highly non-generic. We show  
in the following that it does not apply to a gas of hard spheres, 
neither in the dilute nor in the dense phase.

\noindent
Systems of hard spheres have properties similar to
real fluids and solids and provide a reference 
for successful perturbation theories
\cite{hs}.  Simulations with this model were able to uncover 
fundamental aspects of collective particle dynamics 
such as recollisions  and the ``cage'' effect \cite{hs}.
Hard-sphere systems are also paradigms for the chaotic
and ergodic properties of many-body systems with short
range interactions, and were shown to have a positive
Kolmogorov-Sinai entropy \cite{BDPD,DP}.

\noindent
For dilute gases, Krylov \cite{krylov} provided an
analytical estimate for the maximum Lyapunov exponent,
\begin{equation}
\lambda_1 = - \left(32 \pi K/ 3mN \right)^{1/2} 
\sigma^2 n \log \left( 
\pi \sigma^3 n /\sqrt{2} \right) \quad ,
\label{kry}
\end{equation}
where $K$ is the kinetic energy,
$N$ is the number of particles, $m$ the particle mass, 
$n$ the number denstiy, 
and $\sigma$ is the hard sphere diameter.
This expression has been verified numerically
(apart from a 
factor $\sim 2.8$\cite{DP}), and has been
extended to larger densities \cite{ZON}.

\noindent
   The diffusion coefficient 
for dilute hard-sphere gases is well approximated
by the Enskog expression \cite{CHAP}
\begin{equation}
D_E = \left(3 \pi K/32 mN \right)^{1/2} \frac{1}
{n \pi \sigma^2} \left[1 + \frac{5 n \pi \sigma^3}{12}\right]^{-1} .
\label{dif}
\end{equation}
A comparison of Eqs. (\ref{kry}) and
(\ref{dif}) reveals that, in the
dilute gas limit, the proposed relation $\lambda_1
\propto D^{1/3}$ of Barnett {\it et al.} cannot be satisfied. 
Moreover, we combine in Fig. 1 recent simulation results for $D$
and $\lambda_1$,  which were obtained for a system of 500 hard spheres
over the full range of fluid densities
$(0.0001 < n\sigma^3 < 0.89)$. Reduced units are
used for which $\sigma$, $m$, and the kinetic energy per 
particle $K/N$ are all unity. One observes that
these data are not consistent with the proposed
$D^{1/3}$-dependence (solid line),
neither for low densities nor for large. 

\noindent
     We conclude that the conjecture by Barnett {\it et al.} does
not apply to many-body systems with short-range interactions. But
even its applicability for long-range interactions is doubtful.
A one-dimensional gravitational system with finite $N$ exhibits a
positive $\lambda_1$ \cite{MPT}, whereas this clustering and confining 
system does not show diffusion.  We also note that,
while the theoretical expression  (26) in Ref. \cite{barn1} has been 
obtained for a dilute gas, the data in Fig. 1 of Ref. \cite{barn1} are for
a dense plasma with a Coulomb coupling costant $\Gamma$ ranging from 1 to 150. 
As reported by the same authors \cite{barn2}, for $\Gamma > 1$ the plasma 
behaves as a liquid and not as a gas. The dilute gas limit is
recovered only for $\Gamma \ll 1$ \cite{barn2}.

\noindent
We thank M. Antoni, U. Balucani, A. Rapisarda, and S. Ruffo for 
useful discussions.

\putsmall{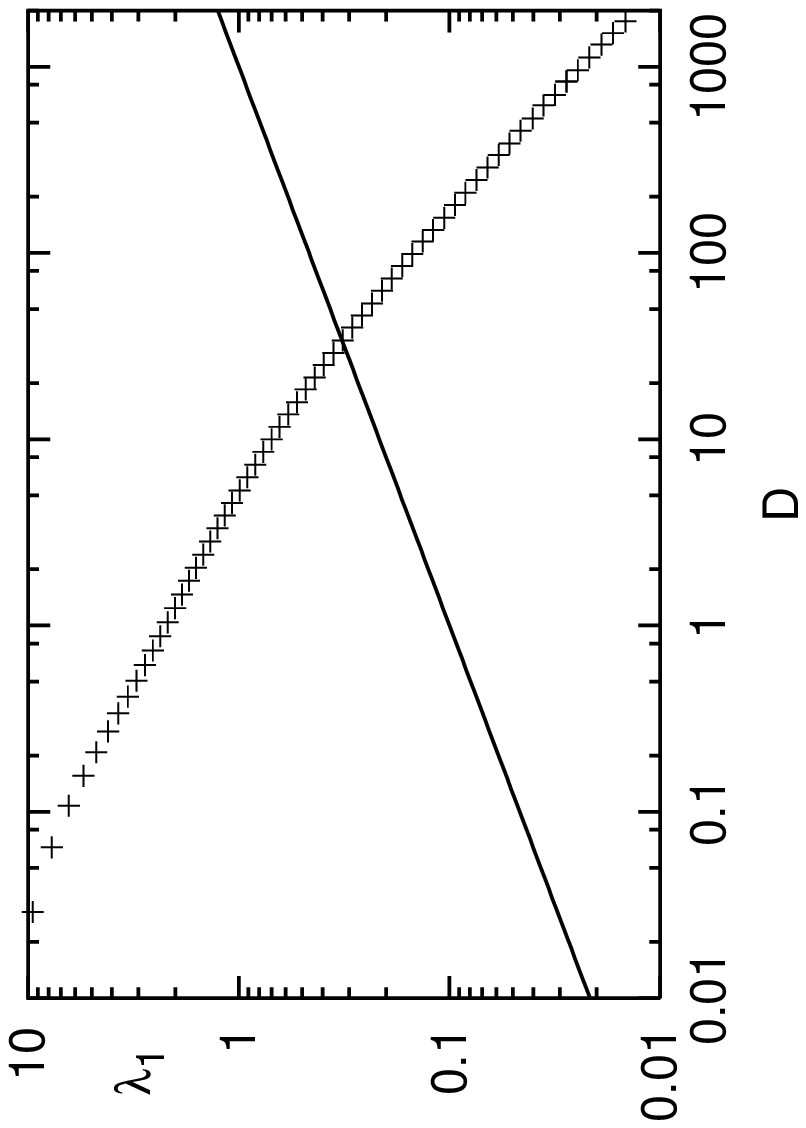}

\begin{figure}
\noindent
{\bf Fig.1}: {\eightrm Simulation results $\lambda_1 = \lambda_1(D)$ (crosses)
for a gas of hard spheres.
The solid line refers to the expression $\lambda_1 \propto  D^{1/3}$ suggested in Ref.
\protect\cite{barn1}
}
\end{figure}

\noindent
A. Torcini$^{(1)}$, Ch. Dellago$^{(2)}$, and  H.A. Posch$^{(3)}$

\noindent
(1) Dipartimento di Energetica, via S. Marta 3, I-50139 Firenze, Italy \\
    INFM, Unit\`a di Firenze, Italy. \\
(2) Department of Chemistry, University of California, Berkeley, 
    CA 94720, U.S.A.\\
(3) Institut f{\"u}r Experimentalphysik, Universit{\"a}t Wien, 
    Boltzmanngasse 5, A-1090 Wien, Austria


\end{document}